\documentclass{article}

\usepackage{arxiv}

\usepackage[utf8]{inputenc} 
\usepackage[T1]{fontenc}    
\usepackage{hyperref}       
\usepackage{url}            
\usepackage{booktabs}       
\usepackage{amsfonts}       
\usepackage{nicefrac}       
\usepackage{microtype}      
\usepackage{lipsum}		
\usepackage{graphicx}
\usepackage{subfig}
\usepackage{natbib}
\usepackage{doi}

\title{An ensemble of data-driven weather prediction models for operational sub-seasonal forecasting}

\author{ 
	Jonathan A. Weyn\thanks{Corresponding author}\hspace{1.5mm}\thanks{Equal contribution. Other authors listed in alphabetical order.} \\
	Microsoft \\
	Redmond, WA \\
	\and
	Divya Kumar\footnotemark[2] \\
	Microsoft \\
	Redmond, WA \\
	\and
	Jeremy Berman \\
	Microsoft \\
	Redmond, WA \\
	\and
	Najeeb Kazmi \\
	Microsoft \\
	Redmond, WA \\
	\and
	Sylwester Klocek \\
	Microsoft \\
	Redmond, WA \\
	\and
	Pete Luferenko \\
	Microsoft \\
	Redmond, WA \\
	\and
	Kit Thambiratnam \\
	Microsoft \\
	Redmond, WA \\
}

\renewcommand{\shorttitle}{Ensemble of AI models for S2S forecasting}

\hypersetup{
pdftitle={An ensemble of data-driven weather prediction models for operational sub-seasonal forecasting},
pdfsubject={ao-ph, cs},
pdfauthor={Jonathan A. Weyn},
pdfkeywords={Weather forecasing, Deep learning, S2S forecasting, Ensemble models},
}

\begin{document}
\maketitle

\begin{abstract}
	We present an operations-ready multi-model ensemble weather forecasting system which uses hybrid data-driven weather prediction models coupled with the European Centre for Medium-range Weather Forecasts (ECMWF) ocean model to predict global weather at 1-degree resolution for 4 weeks of lead time. For predictions of 2-meter temperature, our ensemble on average outperforms the raw ECMWF extended-range ensemble by 4-17\%, depending on the lead time. However, after applying statistical bias corrections, the ECMWF ensemble is about 3\% better at 4 weeks. For other surface parameters, our ensemble is also within a few percentage points of ECMWF's ensemble. We demonstrate that it is possible to achieve near-state-of-the-art subseasonal-to-seasonal forecasts using a multi-model ensembling approach with data-driven weather prediction models.

\end{abstract}


\section{Introduction}

In the last few years, early toy models for data-driven weather prediction \citep{dueben_challenges_2018, weyn_can_2019} have yielded into highly accurate and reliable forecasting systems. Pangu-Weather \citep{bi_accurate_2023} and GraphCast \citep{lam_learning_2023} were the first models to attain the same or better accuracy than traditional numerical weather prediction (NWP) models, such as the European Centre for Medium-range Weather Forecasts (ECMWF) Integrated Forecast System (IFS), on many prediction tasks. Meanwhile, these data-driven models are joined by a rapidly-expanding repertoire of other weather forecasting models (e.g., \cite{keisler_forecasting_2022}, \cite{pathak_fourcastnet_2022}, \cite{chen_fuxi_2023}, \cite{chen_fengwu_2023}, and \cite{nguyen_scaling_2023}).

While the accuracy of these models is impressive, we believe that some important details necessary for the full utilization of these models have not been sufficiently addressed. Some of these limitations include
\begin{enumerate}
	\item Many of the models have not been tested on extended-range forecasting in the two- to six-week subseasonal-to-seasonal (S2S) time scales. Despite the importance of these time scales for a variety of applications such as agriculture and risk management, the models cannot be expected to generalize. While some data-driven models, including the DLWP ensemble of \citep{weyn_sub-seasonal_2021}, FuXi-S2S \citep{chen_fuxi-s2s_2023}, the SFNO \citep{bonev_spherical_2023}, and NeuralGCM \citep{kochkov_neural_2024}, have been run at S2S or longer time scales, some key considerations, including the need for post-processing, that have not been fully addressed.
	\item The models have predominantly been chasing deterministic forecast accuracy, while probabilistic forecasts are often more useful for decision-making. Some papers have developed ensemble models, including DLWP, GenCast \citep{price_gencast_2023}, NeuralGCM, and FuXi-S2S, but all use different methods and have not been compared. Additionally, while some models have been run in operations (c.f., notably, ECMWF's \href{https://github.com/ecmwf-lab/ai-models}{ai-models} repository, and GraphCast's release of a tuned model), questions remain about operationalization of ensemble forecasts, including generalization to initial condition data.
	\item The use of regression losses in training data-driven weather models has resulted in deterministic forecasts which overly smooth fine-scale features, limiting the models' ability to capture the full range of weather phenomena. It also ``games'' the target metrics such as the root-mean-squared-error (RMSE), which are optimized by ensemble-mean-like forecasts, making it difficult to benchmark the actual skill of different models (some models perform particularly well on RMSE at longer lead times because of this smoothing). Notably, models based on denoising diffusion (GenCast) or adverserial losses \citep[Kunyu, ][]{ni_kunyu_2023} have shown promise, even if they may be worse on RMSE.
	\item Data-driven weather forecasting models have until now sought to establish themselves as the most accurate; however, no approach to leverage the strengths of various model architectures or a combination of data-driven and traditional NWP models has been proposed.
\end{enumerate}

In this work, we propose an ensemble of data-driven weather prediction models for operational sub-seasonal forecasting. We aim to partially address some of the above limitations by developing a model that is operationalized, tested on extended-range forecasting, and provides probabilistic forecasts. We also train multiple models based on several architectures to demonstrate the feasibility of a multi-model data-driven approach, which has been successful for traditional NWP \citep[e.g., ][]{kirtman_north_2014}, and propose a combined ensemble of both data-driven and traditional NWP models. The scope of this work is limited to the development of the ensemble and the demonstration of its skill on a few selected tasks. Notably, we do not aim to provide a deep dive into the development of the individual models, or detailed evaluation metrics, but rather to demonstrate the feasibility of the ensemble approach.

\section{Methodology}

\subsection{Data}
All the models are trained on a subset of the ECMWF ERA5 reanalysis dataset \citep{hersbach_era5_2020} conservatively re-gridded to a spatial resolution of 1 degree. Even though our models' temporal resolution is 6 hours, we include data samples at every 3 h. The weather parameters include geopotential (z), temperature (t), specific humidity (q), and zonal and meridional wind components (u and v) at five atmospheric pressure levels (1000, 850, 700, 500, and 200 hPa), as well as the following single-level parameters: 2-meter temperature, 2-meter dewpoint, 10-meter u and v wind components, total column water vapor, mean sea-level pressure, total cloud cover, and sea surface temperature (SST). The training data spans from 1979 to 2014, with validation and testing spanning 2015-2016 and 2017-2018 respectively. Each model takes as input two time steps of data ($t$ and $t-6$ hours) and predicts the weather parameters at the next two time steps ($t+6$ and $t+12$ hours). Forecasts can be made indefinitely into the future by using the predicted values as input for the next time step in an autoregressive fashion.

In addition to the ERA5 input data, we also prescribe the known incoming solar radiation (insolation), topography, and land-sea mask, and SSTs as features at each forward integration step. The insolation provides information about diurnal and seasonal cycles to the model. During training, the SSTs are taken from the ERA5 dataset, while during forecasting, they are taken from the ECMWF IFS model. This SST input is essentially coupling the data-driven models with a physics-based ocean model, which adds key information about long-term variability needed for subseasonal forecasting.

At inference time, the input features are taken from the extended-range (46 days) ensemble forecast system (EEFS) of the ECMWF IFS model, available through the Meteorological Archive and Retrieval System (MARS). Since an upgrade in June 2023, the EEFS provides 100 perturbed members with different initial conditions. Each of our five data-driven models is run on twenty of these initial conditions such that our ensemble size is also 100.

\subsection{Models}

For our ensemble of data-driven weather prediction models, we use five trained models based on three architectures: a model based on MS-Nowcasting \citep{klocek_ms-nowcasting_2022}, which in turn is loosely a convolutional neural network with long short-term memory (ConvLSTM); a transformer backbone with adaptive Fourier neural operators \citep[FourCastNet, ][]{pathak_fourcastnet_2022}; and a graph neural network (GNN) based on GraphCast \citep{lam_learning_2023}. Each model was independently optimized and validated for S2S forecasts and ranked for selection in the ensemble. In brief, the models are as follows:
\begin{itemize}
	\item A modified MS-Nowcasting model. A combination of ConvNeXt \citep{liu_convnet_2022} and ConvLSTM autoencoders with customized dilated convolutional blocks, it has approximately 15 million parameters.
	\item A FourCastNet model. Based on the FourCastNet architecture, we reduced the patch size to only $2\times2$ to reduce the effect of downsampling. It has approximately 46 million parameters.
	\item A GNN model trained with the GraphCast implementation in NVIDIA's \href{https://developer.nvidia.com/modulus}{Modulus} library. It uses a 6-level multi-mesh and has approximately 20 million parameters. Unlike other models, this checkpoint did not undergo the first phase of autoregressive fine-tuning (see Section~\ref{sec:training}).
	\item A GNN model based on our own implementation of GraphCast, which also has a 6-level multi-mesh and approximately 20 million parameters. It differs slightly from the Modulus implementation in optimizer parameters.
	\item Another GNN like the previous model. This one, though, also accepts as input features the insolation and SSTs from the \emph{target} time steps, in addition to the input time steps. 
\end{itemize}

\subsection{\label{sec:training}Training strategy}

Similarly to the autoregressive training strategy in FourCastNet \citep{pathak_fourcastnet_2022} and GraphCast \citep{lam_learning_2023}, we first pretrain the models on the ERA5 dataset using a mean squared error (MSE) loss on just one forward iteration. The learning rate is adjusted using cosine annealing following a linear ramp-up. We then fine-tune the model using a lower fixed learning rate with two forward steps (i.e., four time steps with a maximum lead time of 24 hours), backpropagating the loss through the two steps. This is done to ensure that the model is trained on the same autoregressive policy used for inference, to reduce the likelihood of large errors accumulating over time.

To optimize the model for the operational forecasting task, we finally apply another fine-tuning training using data from the IFS operational analysis from 2017-2022. The smaller dataset and continued use of two autoregressive steps helps ensure the model does not overfit too much to the operational data.

\subsection{Statistical bias correction}

In extended-range forecasting for subseasonal or longer time scales, NWP models are prone to systematic biases such as model drift, which occurs because physical conservation laws are not perfectly followed in weather forecasting models. To correct for this, ECMWF applies in its operational subseasonal-to-seasonal (S2S) forecasts a statistical bias computed using the average known forecast error over hindcasts run for the previous 20 years. For each Monday and Thursday 00 UTC initialization, a hindcast of 10 ensemble members is run for the previous 20 years, and the average error is computed for each lead time averaged over the ensemble members and over each lead week. The average error for three calendar dates (one prior and one after the date of interest) is then used to correct the operational forecast. 

This bias correction has a substantial impact on the performance of the ECMWF baseline. Despite evidence that data-driven models also have systematic biases, no recent literature other than \cite{weyn_sub-seasonal_2021} has attempted to apply a similar bias correction to these models. For this work, we also download all the hindcast initial conditions from MARS to compute the equivalent correction for hindcasts from our ensemble of data-driven models. There is no concern about overlapping the training data with the hindcast data, as the hindcast data is known \emph{a priori} to making the operational forecasts.

\section{Results}

Our goal here is to broadly demonstrate the performance of our ensemble of data-driven weather prediction models. To compare with the latest 100-member ECMWF IFS extended-range ensemble, we limit our test set to forecasts issued on Mondays and Thursdays at 00 UTC time starting on July 3, 2023 until Dec 31, 2023. All verification data are from the IFS operational analysis. We focus on the 2-meter temperature and other surface parameters, as these are crucial for S2S forecasting applications. We compare the performance of our ensemble to the raw ECMWF extended-range ensemble, as well as the ECMWF ensemble after applying the statistical bias correction. We only present aggregated metrics, as a more detailed evaluation is beyond the scope of this work.

\begin{figure}
	\centering
	\includegraphics[width=\textwidth]{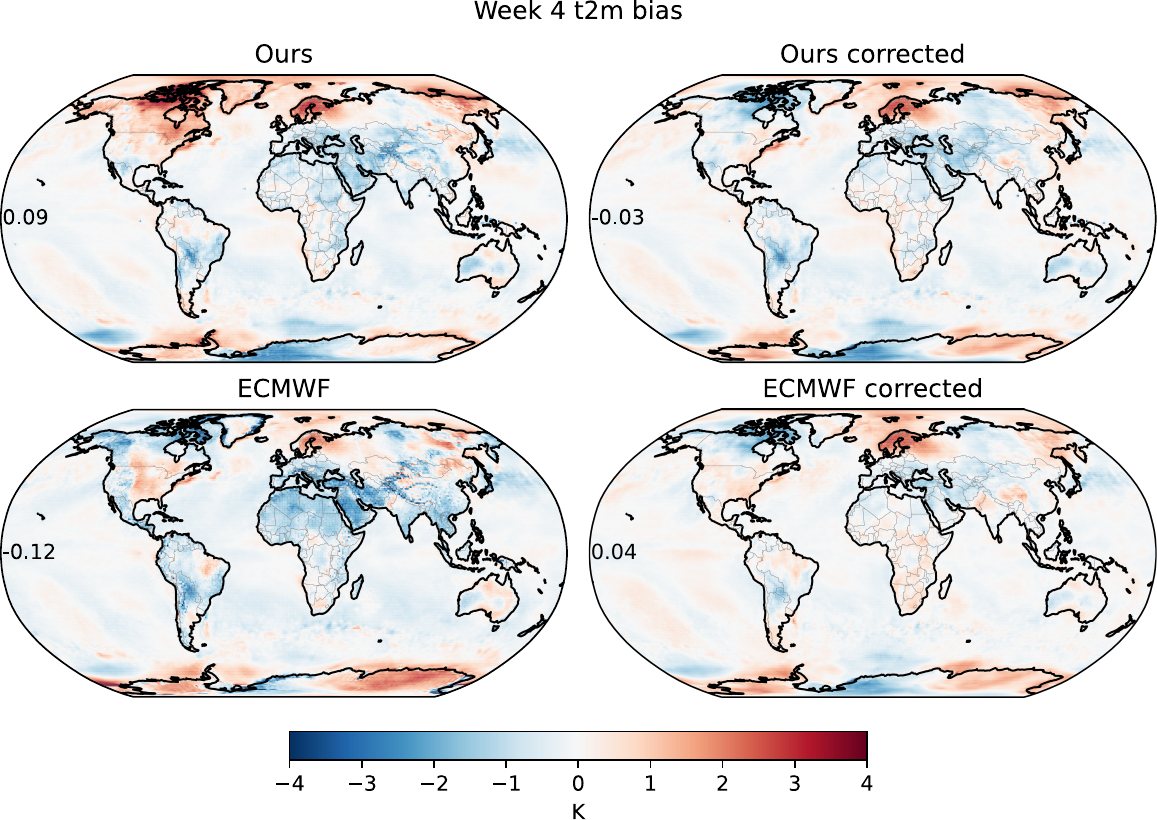}
	\caption{\label{fig:bias_map}\textbf{Spatial distribution of the bias in 2-meter temperature.} The bias is computed as the difference between the ensemble mean of the operational forecast and the ERA5 reanalysis. Numbers in each subplot are the global average.}
\end{figure}

Figure~\ref{fig:bias_map} shows the spatial distribution of the bias in 2-meter temperature for the ECMWF ensemble and our ensemble. The bias is computed as the difference between the ensemble mean of the operational forecast and the operational analysis verification. While our uncorrected model has generally smaller biases compared to the raw ECMWF ensemble (with the exception of the warm bias over Canada), the effect of bias correction is clearly improving the ECMWF results more than those of our ensemble. Nevertheless, bias correction is still helping, particularly in the case of the aforementioned warm Arctic.

\begin{figure}
	\centering
	\subfloat[]{\includegraphics[width=0.45\textwidth]{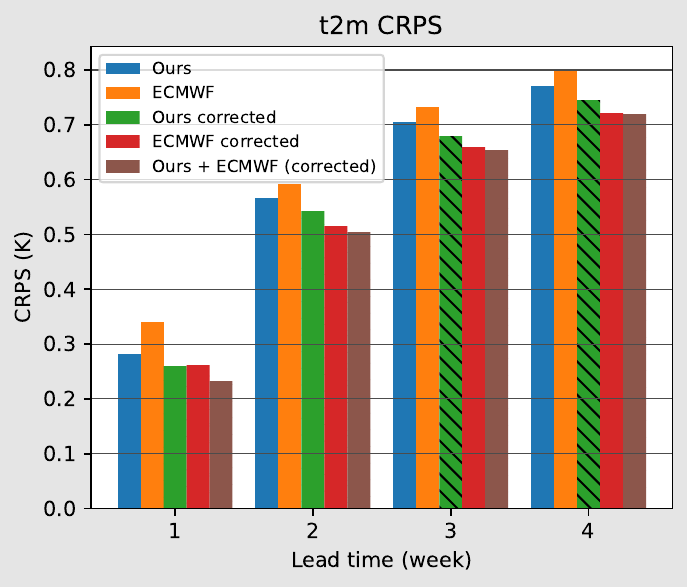}}
	\hfill
	\subfloat[]{\includegraphics[width=0.45\textwidth]{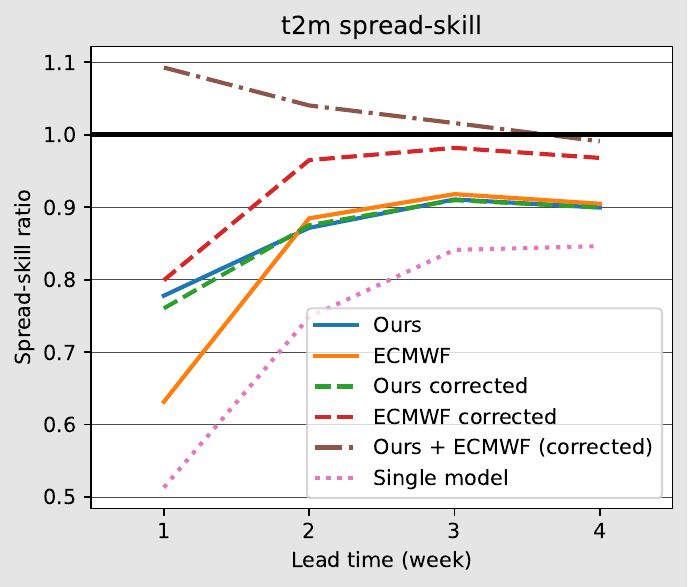}}
	\caption{\label{fig:t2m_skill}\textbf{CRPS and ESSR for 2-meter temperature.} (a) CRPS for 2-meter temperature, as a function of lead time (weeks). (b) Ratio of ensemble spread to skill.}
\end{figure}

We measure the performance of the ensembles with the headline continuous ranked probability score (CRPS) and the ensemble spread-skill ratio (ESSR) metrics. All metrics are area-weighted. The CRPS is a proper scoring rule; that is, it is optimized when the distribution of the ensemble matches the expected distribution of observations. Meanwhile the ESSR is the ratio the ensemble spread (standard deviation) to the skill (root-mean-squared error of the ensemble mean) of the ensemble; it roughly measures whether the variance in the ensemble is well-calibrated. Figure~\ref{fig:t2m_skill} shows the CRPS and ESSR for 2-meter temperature. The CRPS for our ensemble is about 4\% (17\%) lower than that of the raw ECMWF ensemble at week 4 (week 1), indicating better probabilistic forecasts. However, after applying the statistical bias correction, the ECMWF ensemble is about 3\% better at 4 weeks. The modest impact of bias correction on our ensemble's performance may be due to a lower overall bias in the raw predictions, or a lack of consistency between forecasts made on the train versus test sets. Our uncorrected ensemble has a spread-skill ratio of about 0.9, nearly the same as that of the raw ECMWF ensemble. After bias correction, the ECMWF ensemble's spread-skill ratio of increases to nearly 1, indicating a better-calibrated ensemble. Interestingly, if we compare the spread in a 100-member ensemble with only initial condition perturbations (using only the best-performing GNN architecture, shown as the pink dashed line in Fig.~\ref{fig:t2m_skill}b) to that of our multi-model ensemble (blue line), we see that the multi-model approach substantially increases ensemble spread, particularly at early lead times.

Also included in the above results is a combined ensemble using 200 total members, 100 from our ensemble and 100 from the ECMWF ensemble. The effect of combining the data-driven and NWP forecasts clearly boosts the spread in the ensemble, indicated by an ESSR of 1.1 at week 1, and leading to slightly improved CRPS scores at all lead times relative to the corrected ECMWF ensemble. This suggests that the data-driven models are capturing different sources of uncertainty than the NWP models, and that the combination of the two can lead to better probabilistic forecasts.

\begin{figure}
	\centering
	\includegraphics[width=\textwidth]{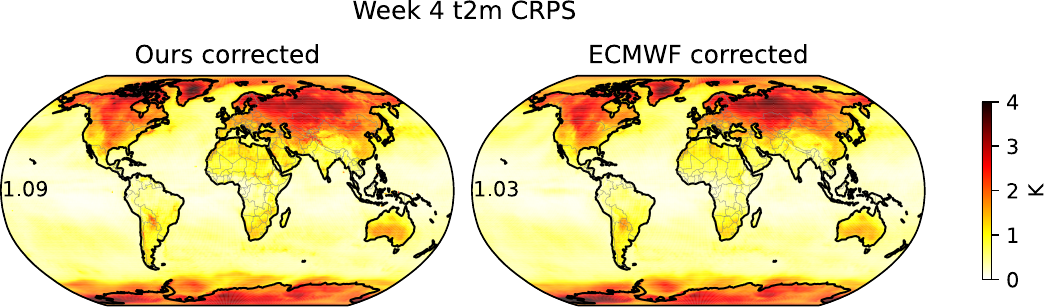}
	\caption{\label{fig:crps_map}\textbf{Spatial distribution of the CRPS score for 2-meter temperature.} The CRPS indicates the quality of the probabilistic forecasts (lower is better) and reduces to the mean absolute error for a deterministic forecast.}
\end{figure}

The spatial distribution of the CRPS score at week 4 for 2-meter temperature is shown in Figure~\ref{fig:crps_map} for both corrected ensembles. Remarkably, the error patterns are nearly identical, suggesting that the data-driven models are capturing the same sources of uncertainty as the NWP models. When evaluating the aggregated metrics over land only, our ensemble increases its lead in week-4 CRPS over the raw ECMWF ensemble to 14\% instead of 4\%, but after corrections the scores remain in favor of ECMWF.

\begin{figure}
	\centering
	\includegraphics[width=0.7\textwidth]{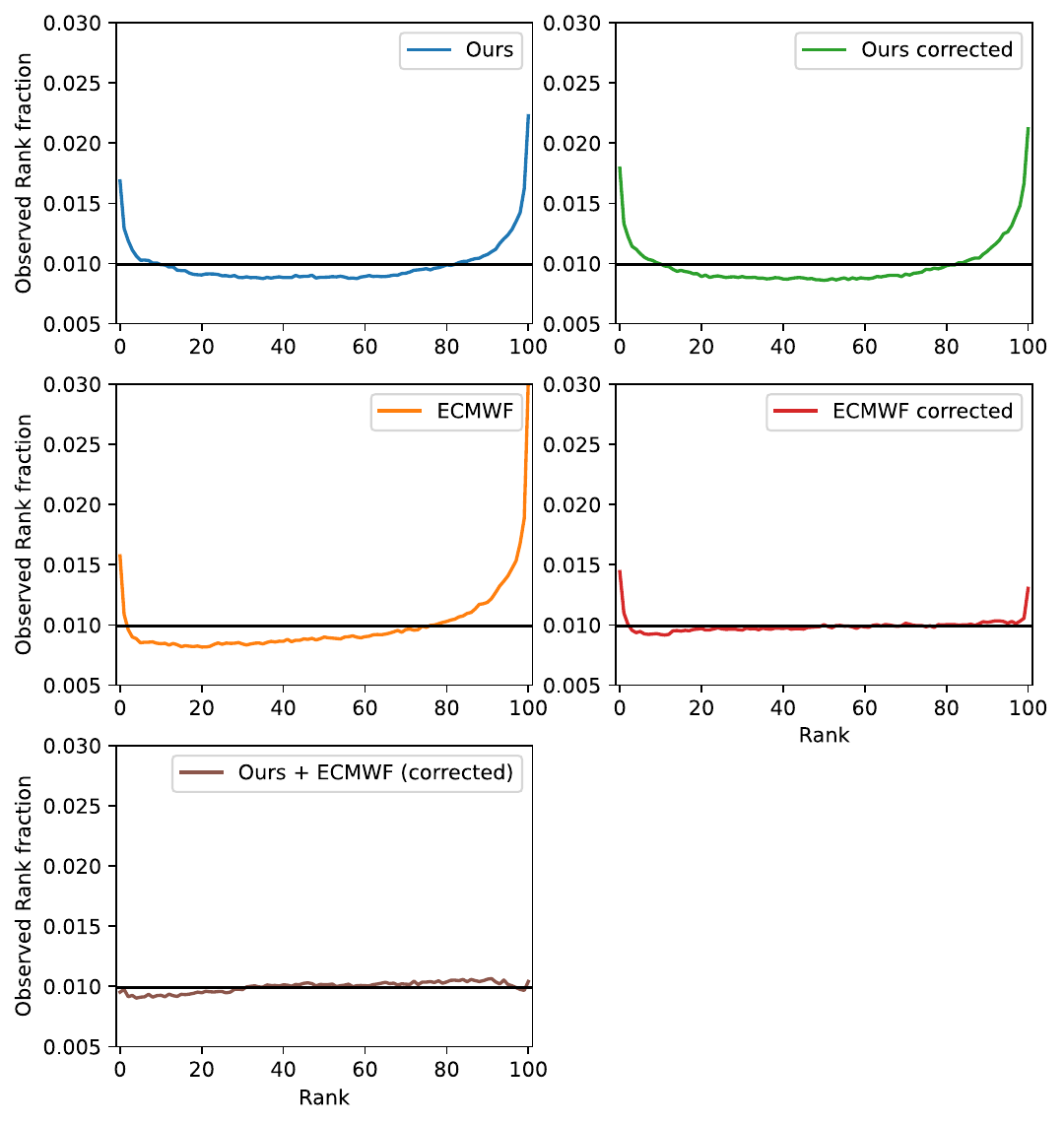}
	\caption{\label{fig:rank_hist}\textbf{Rank histogram for 2-meter temperature for each ensemble.} The rank histogram shows the distribution of the rank of the operational analysis verification within the ensemble.}
\end{figure}

Another indication of ensemble reliability is the rank histogram, which should be approximately uniform if the ensemble is well-calibrated. Figure~\ref{fig:rank_hist} shows the rank histogram for the 2-meter temperature forecasts at week 4. The rank histogram for our ensemble, including after bias corrections, is clearly U-shaped, indicating that many observations fall outside the range predicted by the ensemble, and thus the ensemble is under-dispersive. The rank histogram for the raw ECMWF ensemble is also under-dispersive, and also clearly shows the effect of the overall model cold bias. After applying the statistical bias correction, the rank histogram for the ECMWF ensemble is much improved. The combined ensemble, however, has a rank histogram that is nearly uniform, indicating that the combination of the two ensembles has improved the reliability of the forecasts.

\begin{figure}
	\centering
	\subfloat[]{\includegraphics[width=0.45\textwidth]{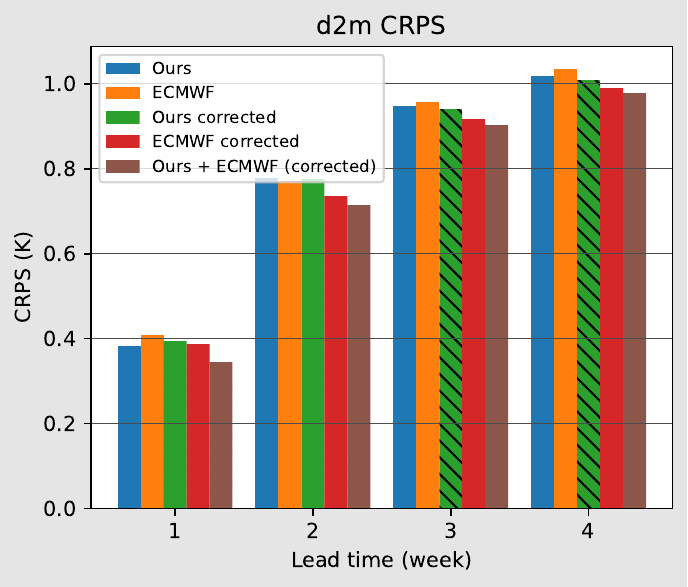}}
	\hfill
	\subfloat[]{\includegraphics[width=0.45\textwidth]{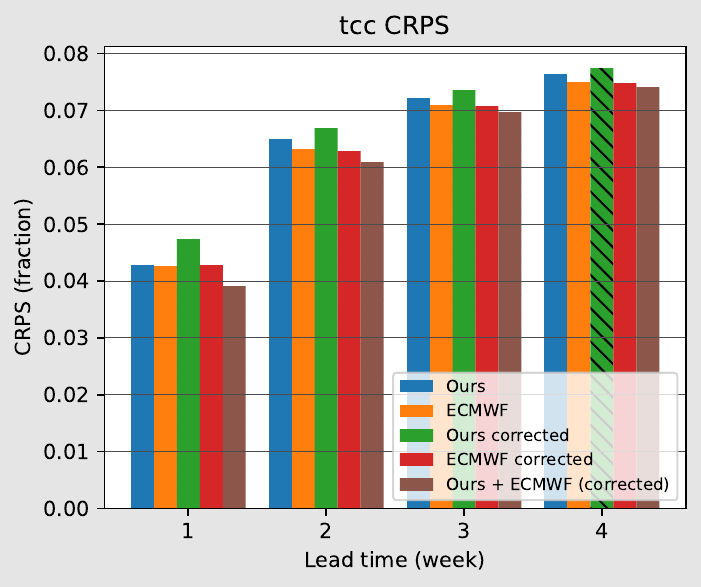}}
	\caption{\label{fig:others_skill}\textbf{CRPS for other surface parameters.} (a) CRPS for 2-meter dewpoint temperature, as a function of lead time (weeks). (b) As in (a) but for total cloud cover.}
\end{figure}

Finally, we also show CRPS scores for 2-meter dewpoint temperature and total cloud cover in Figure~\ref{fig:others_skill}. The CRPS scores for these parameters are within a few percentage points of the ECMWF ensemble, indicating that our ensemble is also competitive for these parameters. The effect of bias correction is reduced and sometimes even detrimental for these parameters. Our ensemble also predicts 6-hourly accumulated precipitation with a diagnostic model similar to that of \cite{pathak_fourcastnet_2022}, but this model scores about 12-25\% worse than the ECMWF ensemble (not shown). We attribute this primarily to inadequacy in the diagnostic precipitation model, which needs to be improved, and in particular, fine-tuned for application on long lead time forecasts.

\section{Discussion}

Using a multi-model ensembling approach, we have shown that near-state-of-the-art sub-seasonal-to-seasonal forecasts are possible using data-driven weather forecasting models. Our ensemble is competitive with the ECMWF extended-range ensemble, and even outperforms the latter by some measures. The combination of the two ensembles, even when ours is weaker, leads to equal or better probabilistic forecasts than either ensemble alone. Like other data-driven weather forecasting models, ours are extremely time-efficient to run, requiring less than 2~hours to produce 100 30-day forecasts on a single NVIDIA V100 graphics processing unit, including all data input/output time. Returning to the key points raised in the Introduction, our work contributes to the growing repertoire of operational data-driven weather forecasting ensemble models in a few key ways:

\begin{enumerate}
	\item Our models have been tested on extended-range forecasting on time scales longer than 15 days, and have shown competitive performance with the latest state-of-the-art in extended-range ensemble forecasting. Notably, we also consider the effect of statistical bias correction on the performance of the models at these time scales.
	\item Like some other papers in the field, we have addressed the need for probabilistic forecasts at longer lead times. We have not, however, attempted to directly compare different ensemble generation methods.  We have shown the feasibility of running an extended-range ensemble of models with operations-ready data, but the hindcast corrections show that there may still be issues with consistency between the training and test sets.
	\item Our models continue to use regression losses, and thus are not an improvement towards better capturing realistic weather patterns at all scales. However, we limit the model fine-tuning to only 24 hours of lead time as a way to limit smoothing of the forecasts, and clearly present ensemble results for fair comparison to other ensemble methods.
	\item Our multi-model approach is yet another way to generate ensembles, but notably, instead of optimizing a single model for best performance, we can leverage the strengths of various model architectures. We also show that a simple combination of data-driven and traditional NWP models can lead to better probabilistic forecasts, particularly by improving the reliability of the ensemble as seen in the rank histogram (Fig.~\ref{fig:rank_hist}).
	
\end{enumerate}

Our results show promise for operationalizing ensembles of data-driven weather forecasting models. However, many open questions still remain for future research. Most notably, the scope of our evaluations was limited and did not consider extreme events such as heat waves and drought, which are critical to forecast at subseasonal time scales. Additionally, we did not address some remaining challenges in the development of the individual models, such as the need for better precipitation forecasts, or other ensemble generations methods such as learned denoising diffusion models. Nevertheless we hope that this work will lead to further efforts to optimize ensemble forecasts by fully leveraging the strengths of different model approaches and a combination of data-driven and traditional NWP methods.

\section*{Acknowledgments}

The authors would like to thank ECMWF for providing all the data used in this work. We also thank Taimoor Akhtar, Haiyu Dong, Richard Turner, Paul Viola, Volodymyr Vragov, and Ruixiong Zhang at Microsoft, along with Matthew Chantry, Mariana Clare, and Zied Ben Bouallegue from ECMWF, for useful discussions and feedback.

\bibliographystyle{unsrtnat}
\bibliography{references}

\end{document}